\documentstyle[12pt]{article}

\newcommand{\diag}{{\rm diag}}
\newcommand{\Ker}{{\rm Ker}\,}
\newcommand{\Mat}{{\rm Mat}\,}

\def\cit#1{\cite{[#1]}}
\def\half{{1 \over 2}}

\def\C{\hbox{\vrule width 0.6pt height 6pt depth 0pt \hskip -3.5pt}C}
\def\R{{I\!\!R}}

\def\Z{Z\!\!\! Z}
\def\im {\imath}
\def\th {\theta}
\def\pa {\partial}
\def\Im {\mbox{Im }}
\def\Re {\mbox{Re }}
\def\Ran {\mbox{Ran}}
\def\rank {\mbox{rank}}

\newcommand{\be}{\begin{equation}}
\newcommand{\ee}{\end{equation}}
\newcommand{\bea}{\begin{eqnarray}}
\newcommand{\eea}{\end{eqnarray}}
\newcommand{\bbc}{C\kern-8.0pt I}
\newcommand{\bbr}{I\!\!R}

\newcommand{\al}{\alpha}
\newcommand{\La}{\Lambda}
\newcommand{\w}{\omega}

\begin{document}

\thispagestyle{empty}
\hfill {}
\vskip 0.7 true cm
\begin{center}
{\Large \bf
Aharonov-Bohm Effect with $\delta$--type Interaction
}
\end{center}
\vskip 1.5 true cm
\begin{center}
{\large L. D\c abrowski$^{1}$ and P. \v S\v tov\'{\i}\v cek $^{2}$}
\end{center}
\begin{center}
{\it $^{1}$S.I.S.S.A., 34014 Trieste, via Beirut 2-4, Italy\\
 $^{2}$Department of Mathematics, Faculty of Nuclear Science, CTU,\\
       Trojanova 13, 120 00 Prague, Czech Republic
}
\end{center}
\vskip 2 true cm
\begin{center}
{\large Abstract}
\end{center}
\bigskip
\noindent
\hspace{.5in}\begin{minipage}{5in}
A quantum particle interacting with a thin solenoid and a magnetic flux
is described by a five-parameter family of Hamilton operators, obtained
via the method of self-adjoint extensions.
One of the parameters, the value of the flux, corresponds to the
Aharonov-Bohm effect;
the other four parameters correspond to the strength of a singular
potential barrier.
The spectrum and eigenstates are computed
and the scattering problem is solved.
\end{minipage}
\vskip 6.5 true cm
\centerline{Ref. SISSA ~171/96/FM}

\newpage
\section{Introduction}

\noindent
The purpose of this paper is to obtain and study the most general family of operators 
which describe the essential features of a quantum mechanical particle under
the joint effect of the electromagnetic potential due to a flux $\phi$
together with the potential barrier supported on the infinite thin
shielded solenoid.

Our initial task is to provide a class of well defined operators
corresponding to the formal differential (plus distributional) expression
$$
- \sum_{\ell =1}^3 (\pa /\partial x_\ell - A(\partial_\ell) )^2 -
\lambda\delta (r) \ ,
$$
where $x_1, x_2, x_3$ are the standard coordinates in $\R^3$,
$A=\im(\phi/2\pi r^2)(-x_2 dx_1+x_1dx_2)$ is a pure gauge potential and
$r=\left( (x_{1})^2+(x_{2})^2\right)^{1/2} $.
For that aim, we first reduce the problem to two dimensions by
making use of translational symmetry with respect to the coordinate $x_3$.
Let $r, \theta$ be, respectively, the radial and angular coordinate in
$\R ^2$,
$0\leq r$ and $ 0\leq \theta \leq 2\pi $.
Set $\alpha=-\phi/2\pi$. We concentrate on the case when
$\alpha\not\in\Z$ and owing to the
gauge symmetry $A'=A+e^{-\im n\theta}de^{\im n\theta}$, $n\in\Z$, we can
restrict ourselves to the case
\be
\alpha\in\, ]0,1[\,.
\ee

The method we adopt in this paper is based on self adjoint extensions
of symmetric operators. From this perspective, we try to combine
two well known cases which
were already extensively discussed in the literature.
The first one, with $\alpha = 0$, corresponds to the so called point
interaction (cf. \cit{AGGH-K}) and  was studied in detail in \cit{AGGH-K0}.
An operator in a one-parameter family is defined on a domain
which is characterized by a linear relation between
certain coefficients which are built up
from the asymptotic behaviour, as $r\to 0$,
of (singular) wave functions. The starting point was a symmetric operator
with a domain formed by wave functions with supports separated from
the origin. The deficiency indices turned out to be (1,1).

The second case, with $\lambda = 0$, corresponds to the pure
Aharonov-Bohm (A-B) potential \cit{AB} and was investigated many times on different
levels (see \cit{Henneberger}, \cit{Rui}, \cit{Berry},
\cit{Nicoleau}, \cit{Stovicek}).
The generalized eigenfunctions are required to belong to
$H^{2,2}_{\mbox{\footnotesize loc}}(\R^2\setminus\{0\})$ and at
the origin the regular condition
\be
\label{limf}
\lim_{r\to 0_+} f(r, \theta ) = 0 \
\ee
is imposed. As the Hamiltonian enjoys rotational symmetry the generalized
eigenfunctions are known and the scattering problem is solved explicitly.

However, it has been known already for some time that when decomposing
the Hilbert space into a direct sum,
\be
\label{decom}
L^2(\R^2,d^2x)\simeq L^2(\R_+,rdr)\otimes L^2([0,2\pi],d\theta)=
\oplus_{m=-\infty}^{\infty}L^2(\R_+,rdr)\otimes\chi_m\, ,
\ee
where
\be
\chi_m (\theta ) = (2\pi )^{-1/2} e^{\im m\theta } \ ,
\ee
then the A-B operator decomposes correspondingly and in the channels
$m=-1$ and $m=0$ the boundary condition (\ref{limf})
is not the most general one.
But since the other boundary conditions admit wave functions
which are singular
at the origin they were usually ruled out.

To our opinion, it makes good
sense to consider the most general case and hence to allow even a sort of
interaction between the two channels. Thus we apply to the pure A-B
Hamiltonian exactly the same procedure which was used in the case of point
interactions. The deficiency indices one obtains this way are (2,2) and
this indicates clearly that the result is not simply a superposition
of the two special cases.

\section{Five-parameter family of Hamilton operators
\newline and their resolvents}

\noindent
In order to get operators which can be
consistently interpreted as describing the physical situation we are
interested in,
we start with the case of pure A-B effect
and introduce the point interaction at $0\in\R ^2$ in the usual way.
Namely, first we consider the restriction
of the self-adjoint pure A-B
operator $H$ to the space of functions with supports
outside of $\{ 0\}$,
obtaining thus a closable symmetric operator.
Then we shall find all possible self-adjoint extensions of its closure
$\tilde H$.

The adjoint ${\tilde H}^*$ is defined as the differential operator
$-(\nabla-A(\nabla))^2$ on the the domain
\be
\psi\in{\cal D}(\tilde H^*)\Longleftrightarrow
\psi\in L^2(\R^2)\cap
H^{2,2}_{\mbox{\footnotesize loc}}(\R^2\setminus\{0\})\ \mbox{ and }\
(\nabla-A(\nabla))^2\psi\in L^2(\R^2) \,.
\ee
On general grounds, $\tilde H$ has equal deficiency indices.
To find the corresponding deficiency spaces we employ the decomposition
(\ref{decom}).
Since the orthogonal projection onto $L^2 (\R_+ ) \otimes \chi_m $
commutes with $\tilde H$ on ${\cal D}(\tilde H)$, we can solve the
eigenvalue problem
\be
\label{adjeigen}
{\tilde H}^* f = k^2 f\,,
\ee
with $k=e^{\im\pi/4}$ and $k=e^{\im 3\pi/4}$,
in each sector $m$ of the angular momentum. Setting
$f(r, \theta ) = g(r) \chi_m (\theta )$,
(\ref{adjeigen}) becomes
\be
\label{adjeigen1}
-(\pa^2 /\partial r^2 + 1/r ~\pa /\partial r +
(m+\alpha)^2/r^2 ) g = k^2 g ,
\ee
which, by the standard substitution $r\to kr$, leads to the Bessel equation.

Next, selecting in the two-dimensional space of solutions the one which
vanishes at the infinity, we arrive at the Henkel functions
\be
\label{base}
g (r) =  H^{(1)}_{\vert m+\alpha\vert}(kr) ~.
\ee
To ensure the integrability we still have to control the
asymptotics as $r\to 0_+$.
The case $\alpha=0$ is known; the $L^2$ solutions exist only in
the sector $m=0$
and thus the the deficiency indices are $(1, 1)$.
Assuming now that $ 0< \phi < 2\pi $ and recalling the asymptotics
\be
\label{assy}
H^{(1)}_{\nu} (z) =
-\frac{2^\nu\im}{\sin\pi\nu ~\Gamma (1\! -\!\nu )} z^{-\nu}
+ \frac{2^{-\nu} \im  e^{-\im \pi \nu}}{\sin\pi\nu ~\Gamma (1\! +\!\nu )}
z^{\nu}
+ {\rm O}(z^{-\nu+2}) \ ,
\ee
as $z\to 0$,
the integrability at $0$ means that $2\vert m+\alpha\vert - 1 < 1$,
which selects precisely two angular momentum sectors: $m =-1$ and $m =0$.
Thus the deficiency indices of $\tilde H$
are $(2, 2)$ and the deficiency space $\cal N_{\im}$
is spanned by $f_{\im}^{1}$ and $f_{\im}^{2}$ given by
\be
\label{basis}
f_{\im}^{1}(r,\theta) = N_1 H^{(1)}_{1-\alpha}(kr) e^{-\im \theta }\,,\
f_{\im}^{2}(r,\theta) = N_2 H^{(1)}_{\alpha}(kr)\,,
\ee
where $k= e^{\im \pi/4}$ ($k=\sqrt{\i}$ with $\Im k > 0$) and
the normalization constants $N_1, N_2$ will be determined later on.

This means that all self-adjoint extensions are in one-to-one
correspondence with the elements of the unitary group $U(2)$ and are
determined by boundary
conditions at the origin. We treat them in detail in the next section.
Thus we get, apart of  $\alpha$ characterizing the magnetic flux,
four additional parameters.

It is now straightforward to determine the domain ${\cal D}(\tilde H)$.
As $\tilde H ={\tilde H}^{**}$ it holds true that
$$
\psi\in{\cal D}(\tilde H)\Longleftrightarrow
\psi\in{\cal D}(\tilde H^*)\ \mbox{ and }\
\langle\psi,\tilde H^*\varphi\rangle=\langle\tilde H^*\psi,\varphi\rangle,
\quad\forall\varphi\in{\cal N}_{\im}+{\cal N}_{-\im}\,.
$$
Consequently we find that
$g(r)\otimes \chi_m(\theta)\in{\cal D}(\tilde H)$, with $m=-1,\ 0$,
if and only if $g\in L^2(\R_+,rdr)\cap
H^{2,2}_{\mbox{\footnotesize loc}}(]0,+\infty[)$,
$(\partial_r^2 + r^{-1}\partial_r + r^{-2}(m+\alpha)^2)g\in L^2(\R_+,rdr)$,
and
\be
\label{limW}
\lim_{r\to 0+} r W(g,h_{\pm})=0\,,
\ee
where $h_{\pm}=H^{(1)}_{|m+\alpha|}(\sqrt{\pm\im}r)$
(with $\Im\sqrt{\pm\im}>0$) and the symbol $W(g,h)$ stands for the
Wronskian,
$$
W(g,h):=\bar g\partial_r h - \bar h\partial_r g \,.
$$

At this point, let us make a short digression and recall a
useful formula comparing
resolvents of two self-adjoint extensions. It is stated in the framework
of Krein's approach to the theory of self-adjoint extensions and was
presented originally in \cit{DG}. Let us consider a general situation when
a closed symmetric operator $X$ is given and $A_0$ is a self-adjoint
extension of $X$, $X\subset A_0=A_0^*\subset X^*$. Assume that the
deficiency indices of $X$ are $(d,d)$, $d<\infty$. Set
\be
\label{defN}
{\cal N}_z:=\Ker(X^*-z),\quad R^0_z:=(A_0-z)^{-1},\quad
z\in \C \setminus\R \,.
\ee
The following facts are well known and easy to check. First,
\be
\psi\in{\cal N}_w\Longrightarrow \psi+(z-w)R^0_z\psi =
(A_0-w)R^0_z\psi\in{\cal N}_z\,.
\ee
Second, if $U^0_z:{\cal N}_z\to{\cal N}_{\bar z}$,
$z\in \C \setminus\R$, is the unitary mapping defining the self-adjoint
extension $A_0$ then
\be
\forall\psi\in{\cal N}_z,\quad
U^0_z\psi=-(A_0-z)(A_0-{\bar z})^{-1}\psi \,.
\ee
Fix $w\in \C\setminus\R$ and a basis $\{ f^\ell_w\}_\ell$ in ${\cal N}_w$.
Set
\be
f^\ell_z=f^\ell_w +(z-w)R^0_z f^\ell_w=(A_0-w)R^0_z f^\ell_w,\quad
z\in \C \setminus\R \,.
\ee
Since $(A_0-w)R^0_z$ is injective and
$\dim {\cal N}_w=\dim {\cal N}_z=d$, $\{f^1_z,\dots,f^d_z\}$ is a basis in
${\cal N}_z$. Suppressing the superscript $\ell$ one can verify readily that
\bea
& & \forall z,z'\in \C\setminus\R,\quad
f_z=f_{z'} +(z-z')R^0_z f_{z'}\,, \\
& & \forall z\in \C\setminus\R,\quad U^0_z f_z=-f_{\bar z} \,.
\eea
Thus, in order to reproduce the vector-valued function $f_z$,
one can take
any $z'\in \C\setminus\R$ instead of $w$. Furthermore, the matrix of
$U^0_z$ in the bases $\{ f^\ell_z\}_\ell$, $\{ f^\ell_{\bar z}\}_\ell$
equals $-1$.

To proceed further let us introduce a matrix $P(z,w)=(P^{jk}(z,w))$
of scalar products relating two spaces ${\cal N}_z$ and ${\cal N}_w$,
\be
P^{jk}(z,w):=\langle f^j_z,f^k_w\rangle \,.
\ee
One finds that
\be
P(z,z)=P({\bar z},{\bar z})\,,
\ee
and if $U$ is any matrix expressed in the bases $\{ f^\ell_z\}$,
$\{ f^\ell_{\bar z}\}$, and representing a unitary mapping
${\cal N}_z\to{\cal N}_{\bar z}$  then
\be
P(z,z)=U^* P({\bar z},{\bar z}) U \,.
\ee
Furthermore, if ${\cal P}_z$ is the orthogonal projector onto ${\cal N}_z$
then
\be
{\cal P}_z=\sum_{jk} \Bigl(P(z,z)^{-1}\Bigr)^{jk}f^j_z\,
\langle f^k_z,\cdot \rangle \,.
\ee
Our primary interest is to compare $A_0$ with another self-adjoint
extension $A$ of $X$ corresponding to a family of unitary mappings
$U_z:{\cal N}_z\to{\cal N}_{\bar z}$. Krein's formula tells us that
\be
R_z-R^0_z=({\bar z}-z)^{-1}{\cal P}_z^*(U_{\bar z}-U_{\bar z}^0)
{\cal P}_{\bar z} \,,
\ee
with the symbol ${\cal P}_z^*$ standing for the embedding of ${\cal N}_z$
into the global Hilbert space. This means that there exists a family of
$d\times d$ matrices, $M_z=(M^{jk}_z)$, such that
\be
R_z-R^0_z=\sum_{jk} M^{jk}_z\, f^j_z\,
\langle f^k_{\bar z},\cdot \rangle \,.
\ee
We claim that
\bea
& & M_z-M_w=(z-w)M_z P({\bar z},w)\,M_w, \\
& & ({\bar w}-w)M_w=(U^{-1}+1)P(w,w)^{-1} \,,
\eea
where $U$ is the matrix of $U_w:{\cal N}_w\to{\cal N}_{\bar w}$ in the
bases
$\{ f^j_w\}$ and $\{ f^j_{\bar w}\}$. The proof is quite straightforward
and relies on Krein's formula, the first resolvent identity and the
explicit expression for ${\cal P}_z$. Provided $M_w$ and $M_z$ are
invertible one can also write
\be
M_w^{-1}-M_z^{-1}=(z-w)P({\bar z},w) \,.
\ee
What we shall need in the sequel is the particular choice of $w=-\im$.
In this case,
\be
\label{Mzinv}
M^{-1}_z = 2\im P(\im , \im ) (U+1)^{-1} - (z+\im ) P(\bar z, -\im )\,,
\ee
where this time
$U$ is the matrix of $U_{\im}:{\cal N}_{\im}\to{\cal N}_{-\im}$
in the above specified bases.

Next we apply this general procedure to our problem, with $A_0\equiv H$ --
the pure A-B operator. Thus,
for $z\in \C \setminus\R_+$, we choose in ${\cal N}_{z}$ a particular basis
which depends holomorphically on $z$ by
\be
\label{zbas}
f_{z}^{\ell}= \bigl( 1+(z-\im ) R_z \bigr) f_{\im}^{\ell}, ~~ \ell =1, 2,
\ee
where $R_z$ is the resolvent of the pure A-B operator
defined by its integral kernel (Green function)
\be
\label{ABres}
G_z(r, \th; r', \th ') =
{1\over 2\pi} \sum_{m\in \Z} e^{\im m(\theta - \th ')}
\int_0^\infty (k^2 -z)^{-1} J_{\vert m+\alpha\vert}(kr)
J_{\vert m+\alpha\vert}
(kr') k dk ~.
\ee
Recalling that
\be
\label{HverK}
H^{(1)}_{\nu} (z) = {2\over\pi\im}\, e^{-\im \pi\nu/2} K_{\nu}
(-\im z)\,,
\ee
where
\be
K_{\nu} (z)= \int_0^\infty e^{-z \cosh t}\cosh \nu t ~dt\,,
\ee
and using the formulae 8.13(3) and 8.5(12) of \cit{Bateman},
\bea
\label{bat}
& & \int_0^\infty K_\mu(ay)\, J_\mu(xy)\,y\,dy=\left({x\over a}\right)^\mu
(x^2+a^2)^{-1} \,,\\
& & \int_0^\infty {x^\mu \over x^2+a^2 }\,J_\mu(yx)\,x\,dx=
a^\mu K_\mu(ay),\quad \mu\in\,]-1,3/2[,\ \Re a>0,
\eea
we have
\bea
f_{z}^{1}(r,\theta ) &=& N_1 e^{-\im\pi(1-\alpha )/4}
    (\sqrt{z})^{1-\alpha } H^{(1)}_{1-\alpha}(\sqrt{z}r)
e^{-\im \theta }\,,\nonumber\\
\label{zbase}
f_{z}^{2}(r,\theta ) &=& N_2 e^{-\im\pi\alpha/4}
    (\sqrt{z})^{\alpha} H^{(1)}_{\alpha}(\sqrt{z}r) ~,
\eea
where $z\in\C \setminus\R_+,\ \Im \sqrt{z}>0$. Particularly,
\be
\label{fpart}
f_{-\im}^{1}(r,\theta ) = N_1 e^{\im\pi(1-\alpha )/2}
    H^{(1)}_{1-\alpha}(kr)
e^{-\im \theta } \,,\
f_{-\im}^{2}(r,\theta ) = N_2 e^{\im\pi\alpha/2}
    H^{(1)}_{\alpha}(kr),\ \ee
where $k=\sqrt{-\im}=e^{\im 3\pi/4}$.

A word of warning should be said here. We use the branch
$(e^{\im\varphi})^\nu = e^{\im\varphi\nu}$, for $0\leq\varphi <2\pi$,
which differs from the usual choice made in surveys like
\cit{Grad-Ryzh}, \cit{Bateman}.

Next we compute the matrix $P(z, z')$,
\be
P^{\ell , m}(z, z') :=
\langle f_{z}^{\ell} ,  f_{z'}^{m} \rangle
= P^{\ell , \ell}(z, z') \delta_{\ell , m}, ~~~\ell, m = 1, 2 ~.
\ee
Since $\overline{K_{\nu} (z)} = K_{\nu} (\bar{z})$ and using the formula
6.521 of \cit{Grad-Ryzh},
\be
\int_0^\infty K_\nu(ar)\,K_\nu(br)\,r\,dr=
{\pi(ab)^{-\nu}(a^{2\nu}-b^{2\nu}) \over
2\sin(\pi\nu)(a^2-b^2) } \,,
\ee
we find that
\be
P^{\ell , \ell}(z, z') =
N_{\ell}^2 ~
\frac{4 e^{-\im\pi\nu}}{\sin (\pi\nu)} ~
\frac{(\bar z)^\nu - (z')^\nu}{z - z'},
\ee
where
$$
\nu = \left\{
        \begin{array}{ll}
        1-\alpha &~\mbox{if ~~$\ell = 1$}\\
        \alpha   &~\mbox{if ~~$\ell = 2$}
        \end{array}
    \right. \ .
$$
Making use of the identity
\be
\int_0^\infty |H^{(1)}_\nu(kr)|^2\,r\,dr=
\Bigl(\pi\,\cos(\pi\nu/2)\Bigr)^{-1},\quad k=\sqrt{\pm\im},\
\Im k>0,
\ee
we choose now
\be
N_1 = 2^{-1/2} \sin^{1/2}(\pi\alpha /2),\quad
N_2 = 2^{-1/2} \cos^{1/2}(\pi\alpha /2)\,.
\ee
In this case the basis $\{f_{\im}^{\ell}\}$ (\ref{basis}), as well as
the basis $\{f_{-\im}^{~\ell}\}$,
 is orthonormal, i.e., in the matrix notation,
\be
\label{pii}
P(\im , \im ) = P(-\im , -\im ) = I~,
\ee
where $I$ is the $2\times 2$ unit matrix.
Moreover, introducing the matrix
\be
\label{defD}
D = \diag (1\! -\! \alpha , \alpha )\,,
\ee
we also have
\be
\label{pzi}
P(\bar z, -\im ) =
(z+\im )^{-1} \sin^{-1} (\pi D/2) e^{-\im\pi D} (e^{3\im\pi D/2} - z^D) ~.
\ee
We conclude that all self-adjoint extensions $H^U$ of $\tilde H$
are bijectively labelled
by $2\times 2$ unitary matrices $U$ by means of
\be
\label{domHU}
{\cal D}(H^{U}) := \{\psi;\ \psi =
\tilde \psi + \sum_\ell c_\ell \psi^{\ell}\} ~,
\ee
where $\tilde \psi \in D(\tilde H)$ and
$$
\label{pbasis}
\psi^{\ell} = f_{\im}^{\ell} + \sum_m f_{-\im}^{m} U^{m, \ell},  ~~
\ell = 1, 2,
$$
and
\be
H^{U} \psi = \tilde H \tilde\psi + \sum_\ell c_\ell
(\im f_{\im}^{\ell} -\im \sum_m f_{-\im}^{m} U^{m, \ell}) ~.
\ee
According to the above discussion, $U = -1$
corresponds to the pure A-B operator $H$.
Moreover, diagonal $U$ describe the extensions preserving the angular momentum
(which otherwise is non-conserved).

\section{Boundary conditions}

\noindent
The family of operators $H^{U}$, defined so far abstractly,
can be equivalently characterized
as differential operators with some well specified
boundary conditions. For this purpose, we introduce
four linear functionals $\Phi_{a}^n$, $n=1, 2$, $a=1, 2$,
corresponding to the two critical angular sectors and to the first two
leading terms giving
the asymptotic behaviour of the radial part of $\psi$ as $r\to 0$.
We define
\bea
\Phi_{1}^1 (\psi ) &:=& \lim_{r\to 0} ~r^{1-\alpha}
  \int_0^{2\pi} \psi (r, \th )\, e^{\im \th}\, d\th /2\pi \,,
  \nonumber\\
\label{funct}
\Phi_{2}^1 (\psi ) &:=& \lim_{r\to 0} ~r^{-1+\alpha}
  \left[ \int_0^{2\pi} \psi (r, \th )\, e^{\im\th}\, d\th /2\pi
            -r^{-1+\alpha}\Phi_{1}^1 (\psi ) \right]\,, \\
\Phi_{1}^2 (\psi ) &:=& \lim_{r\to 0} ~r^{\alpha}
  \int_0^{2\pi} \psi (r, \th )\, d\th /2\pi \,,
  \nonumber\\
\Phi_{2}^2 (\psi ) &:=& \lim_{r\to 0} ~r^{-\alpha}
  \left[ \int_0^{2\pi} \psi (r, \th )\, d\th /2\pi
            -r^{-\alpha}\Phi_{1}^2 (\psi ) \right] \,.\nonumber
\eea
This definition is, of course, dictated by the asymptotic behaviour of
Hankels functions (cf. (\ref{assy})). So for $\psi\in{\cal D}(H^U)$,
the part of $\psi$ which is singular or becomes singular after
differentiation by $\partial_r$ is given by
$$
\Bigl(\Phi_{1}^1(\psi)\,r^{-1+\alpha}+\Phi_{2}^1(\psi)\,r^{1-\alpha}\Bigr)
\,e^{-\im\th}+\Phi_{1}^2(\psi)\,r^{-\alpha}+\Phi_{2}^2(\psi)\,r^{\alpha}\,.
$$

Let us first check the symmetry condition
$\langle\varphi_1,H^U\varphi_2\rangle=
\langle H^U\varphi_1,\varphi_2\rangle$, for
$\varphi_1,\varphi_2\in{\cal D}(H^U)$. The integration by parts gives
($W=$ Wronskian)
$$
\lim_{r\to 0+}\int_0^{2\pi}r\,W(\varphi_1,\varphi_2)\,d\theta=0\,.
$$
Only the singular parts of $\varphi_1,\varphi_2$ can contribute and thus
one arrives at
\be
\label{symmFDF}
\Phi_1(\varphi_1)^*D\Phi_2(\varphi_2)=
\Phi_1(\varphi_2)^*D\Phi_2(\varphi_1),\quad\forall
\varphi_1,\varphi_2\in{\cal D}(H^U)\,,
\ee
where we have introduced
\be
\Phi_a(\psi):=\pmatrix{\Phi_a^1(\psi)\cr \Phi_a^2(\psi)},\quad
a=1,2\,,
\ee
and $D$ was defined in (\ref{defD}).

Next we apply the functionals $\Phi_a^n$ to the functions
$f^{\ell}_{\pm\im}$. Namely,
introduce four matrices $\Phi_{ab}$, where the label $a = 1, 2$
refers to the first (respectively the second) leading coefficient,
and $b= \pm$ refers to $\pm$ in ${\cal N}_{\pm\im}$.
They are defined by
\be
(\Phi_{ab})^{n\ell} := \Phi^{n}_{a} (f_{b\im}^{\ell}) ~;
\ee
so the rows of these matrices are numbered by the angular momentum $n$
(for the sake of convenience we shifted the index by 2,
$n=m+2\in\{1,2\}$)
and the columns by $\ell$, corresponding to the basis in
${\cal N}_{\pm\im}$.
In view of the asymptotic expansion of the functions $f^\ell_{b\im}$
(cf. (\ref{basis}) and (\ref{fpart})),
they read explicitly
\bea
\Phi_{1,+} \! &=&\! \Phi_{1,-}   \nonumber\\
&=& \frac{-\im 2^{-1/2}}{\sin(\pi\alpha)}\,\cos^{1/2}(\pi D/2)\,
    2^D\, \Gamma(1-D)^{-1}\, \exp(-\im\pi D/4) \,,
    \nonumber\\
\label{Fipm}
\Phi_{2,+} \! &=& \!
    \frac{\im 2^{-1/2}}{\sin(\pi\alpha)}\,\cos^{1/2}(\pi D/2)\,
    2^{-D}\, \Gamma(1+D)^{-1}\, \exp(-\im 3\pi D/4) \,, \\
\Phi_{2,-} \! &=&\!
    \frac{\im 2^{-1/2}}{\sin(\pi\alpha)}\,\cos^{1/2}(\pi D/2)\,
    2^{-D}\, \Gamma(1+D)^{-1}\, \exp(\im \pi D/4) \,.\nonumber
\eea
Here and everywhere in what follows we use the obvious notation: if
a function $F$ is well defined on the set $\{1-\alpha,\alpha\}$ then
$$
F(D):=\diag\Bigl(F(1-\alpha),F(\alpha)\Bigr)\,.
$$

   From the definition (\ref{domHU}) of ${\cal D}(H^U)$ follows that
$\psi\in L^2(\R^2)$ belongs to the domain of $H^U$ if and only if
\be
\pmatrix{\Phi_1(\psi) \cr \Phi_2(\psi)}\in\Ran
\pmatrix{\Phi_{1,+}+\Phi_{1,-}U \cr \Phi_{2,+}+\Phi_{2,-}U} \,.
\ee
Inversely, assume that we are given a couple of $2\times 2$ matrices
$X_1, X_2$ such that $\rank(X_1^t,X_2^t)=2$, and consider the boundary
condition
\be
\label{FiRan}
\pmatrix{\Phi_1(\psi) \cr \Phi_2(\psi)}\in\Ran
\pmatrix{X_1 \cr X_2} \,.
\ee
The symmetry property (\ref{symmFDF}) leads to the requirement
\be
\label{symmXDX}
X_1^*D X_2= X_2^*D X_1 \,.
\ee
In fact, relying on the explicit form of the matrices $\Phi_{ab}$, one
can show quite straightforwardly that for any couple $X_1, X_2$ with
the above properties there exist exactly one $2\times 2$ unitary matrix
$U$ and $Y\in GL(2,\C)$ such that
\be
\label{X12Y}
X_1Y=\Phi_{1,+}+\Phi_{1,-}U,\quad X_2Y=\Phi_{2,+}+\Phi_{2,-}U \,.
\ee
On the contrary, if $U$ is unitary then $X_a=\Phi_{a,+}+\Phi_{a,-}U$,
$a=1,2$, verify (\ref{symmXDX}) and $\rank(X_1^t,X_2^t)=2$.
This way we have rederived a well known result               
that all self-adjoint extensions of $\tilde H$
are in one-to-one correspondence with points of
a real 4-dimensional submanifold of the Grassmann manifold $G_2(\C^4)$
determined by the equation (\ref{symmXDX}).

One can rewrite the boundary condition (\ref{FiRan})
in a more convenient form
when making use of the biholomorphic diffeomorphism
$G_2(\C^4)\to G_2(\C^4)^\sharp $. Here $G_2(\C^4)^\sharp$
stands for the Grassmann manifold in the space dual to
$\ \C^4$. The points of $G_2(\C^4)^\sharp$ are
represented by couples of matrices $A_1,A_2\in\Mat(2,\C)$, with
$\rank(A_1,A_2)=2$, modulo the left action of $GL(2,\C)$. The
diffeomorphism is given by
\bea
& & \Ran\pmatrix{X_1 \cr X_2}\in G_2(\C^4)\mapsto
\Ran\pmatrix{A_1^t \cr A_2^t}\in G_2(\C^4)\,, \nonumber \\
\label{grass}
& & \mbox{where }\ A_1X_1+A_2X_2=0 \,.
\eea
The real submanifold of $G_2(\C^4)$ determined by (\ref{symmXDX})
is mapped bijectively
onto the real 4-dimensional submanifold of $G_2(\C^4)^\sharp $ determined
by
\be
\label{ADA}
A_1D^{-1}A_2^*=A_2D^{-1}A_1^* \,.
\ee
We conclude that each self-adjoint extension of $\tilde H$ is determined
by a boundary condition of the type
\be
\label{AFi}
A_1\Phi_1(\psi)+A_2\Phi_2(\psi)=0 \,,
\ee
where $A_1,A_2\in\Mat(2,\C)$ verify (\ref{ADA})
and $\rank(A_1,A_2)=2$. Two couples
$(A_1,A_2)$ and $(A'_1,A'_2)$ define the same self-adjoint operator if
and only if there exists $Y\in GL(2,\C)$ such that $A'_1=YA_1$,
$A'_2=YA_2$.

Let us now restrict ourselves to an open dense subset of the manifold of
boundary conditions (\ref{AFi})
which we obtain by fixing $A_1=I$ and setting
$A_2=-\Lambda$. The condition (\ref{AFi}) means, of course, that
\be
\label{BC}
\pmatrix{
\Phi_{1}^1 (\psi )\cr
\Phi_{1}^2 (\psi )}
=
\La \pmatrix{
\Phi_{2}^1 (\psi )\cr
\Phi_{2}^2 (\psi )} ~,\quad\quad\forall \psi\in {\cal D}(H^U) ~.
\ee
The restriction (\ref{ADA}) then reads
\be
\label{rc}
D \La = \La^* D ~.
\ee
All matrices $\La$ obeying (\ref{rc})
can be parameterized by the aid of four real
(or two real and one complex) parameters, namely
\be
\label{Lambda}
\La = \pmatrix{     u&\alpha\bar w \cr
        (1\! -\!\alpha )w&v } ~,\qquad
\mbox{with } u,v \in \bbr \mbox{ and } w \in \bbc \,.
\ee

  From (\ref{grass}) follows that one can choose $X_1=\La,\ X_2=I$. 
In virtue of (\ref{X12Y}), this leads to a relation between $\La$ and $U$,
\be
\label{LU}
\La = (\Phi_{1,+}+\Phi_{1,-} U)(\Phi_{2,+}+\Phi_{2,-}U)^{-1} ~,
\ee
provided the relevant matrix is invertible.
We use the following parameterization of $U$,
\be
U = e^{\im\w}\pmatrix{ qe^{\im a} & -(1\! -\! q^2)^{1/2}e^{-\im b} \cr
            (1\! -\! q^2)^{1/2}e^{\im b}  & qe^{-\im a} } \,,
\ee
where $a,b,q,\w\in\R$, $0\leq q\leq 1$. The matrix
$(\Phi_{2,+}+\Phi_{2,-}U)$ is invertible exactly when $d \neq 0$ where
\be
\label{loc}
d := \sin\w + q \sin(a\! -\!\pi\alpha) \,.
\ee
In this case, the parameters of the matrix $\La$ can be expressed explicitly in terms of
 $a,b,q,\w$,
\bea
u &=& d^{-1}~ 2^{2\! -\! 2\alpha}
\frac{\Gamma (2\! -\! \alpha )}{\Gamma (\alpha)}
\left(
      \cos\Bigl(\w\! +\! \frac{\pi}{2}\alpha\Bigr)\! +
      \! q \cos\Bigl(a\! -\!\frac{\pi}{2}\alpha\Bigr)
\right) \,,
\nonumber\\
\label{LU1}
\bar w &=& d^{-1}~ 
\left(
      2(1\! -\! q^2) \sin(\pi\alpha )
\right)^{1/2}
e^{-\im (b\! -\!\frac{\pi}{2}\alpha\! -\!\frac{\pi}{4})} \,,
\\
w &=& d^{-1}~ 
\left(
      2(1\! -\! q^2) \sin(\pi\alpha )
\right)^{1/2}
e^{\im (b\! -\!\frac{\pi}{2}\alpha\! -\!\frac{\pi}{4})} \,,
\nonumber\\
v &=& -d^{-1}~ 2^{2\alpha} ~
\frac{\Gamma (1\! +\! \alpha )}{\Gamma (1\! -\! \alpha )}
\left(
      \sin\Bigl(\w\! -\! \frac{\pi}{2}\alpha\Bigr)\! +
      \! q \sin\Bigl(a\! -\!\frac{\pi}{2}\alpha\Bigr)
\right) ~. \nonumber
\eea
Obviously, $u=w=v=0$ corresponds to the pure A-B case.
Moreover, diagonal $\Lambda$ describe the operators preserving the angular momentum
($w$ is responsible for its non-conservation).

\section{Spectrum and eigenspaces}

\noindent
In order to find the spectrum one can use Krein's formula
for the resolvent $R_z^U=(H^U-z)^{-1}$,
\be
R^{U}_z = R_z + \sum_{k\ell} f^{k}_z M^{k\ell}_z
\langle f_{z}^{\ell} , \cdot \rangle ~.
\ee
Using (\ref{pii}), (\ref{pzi}) and (\ref{Mzinv}) we get
\be
M^{-1}_z = \sin^{-1}(\pi D/2)
    \left(z^D e^{-\im\pi D} (U+1) - e^{\im\pi D/2}U -
        e^{-\im\pi D/2} \right) (U+1)^{-1} ~.
\ee
Since $R_z^U$ is a rank two perturbation of $R_z$, $H^U$ and $H$ have the
same absolutely continuous spectrum, namely $[0,\infty[$.

The discrete spectrum is determined by the equation
$\det M_z^{-1}=0$, i.e.,
\be
\label{detu}
\det \left( p^{2D}(U+1) - e^{\im\pi D/2}U -e^{-\im\pi D/2} \right) =0 ~,
\ee
where we have introduced $p>0$ by $p^2 = -z$.
According to the discussion below there are no non-negative eigenvalues.
If $d\not=0$ (cf. (\ref{loc}))
then $(e^{\im\pi D/2}U-e^{-\im\pi D/2})$ is
invertible and owing to (\ref{LU}),
\bea
(U+1)\Bigl(e^{\im\pi D/2}U-e^{-\im\pi D/2}\Bigr)^{-1}=
& - & \cos^{-1/2}(\pi D/2)\, 2^{-D}\, \Gamma(1-D)e^{\im\pi D/4}\, \La
\nonumber\\
& \times & \cos^{1/2}(\pi D/2)\, 2^{-D}\, \Gamma(1+D)e^{-\im\pi D/4} \,.
\eea
In this case, (\ref{detu}) is equivalent to
\be
\label{detL}
\det\left( \frac{\Gamma(1\! -\! D)}{\Gamma(1\!+\! D)}\,
           \left({p\over 2}\right)^{2D} \La +1
    \right) =0 \,.
\ee
Rewriting (\ref{detL})
in terms of the parameters $u, v, w$ (cf. (\ref{Lambda})) we get
\be
\label{speLa}
  \left(
\left({p\over2}\right)^{2\alpha\! -\! 2} +
\frac{\Gamma (\alpha)}{\Gamma(2\! -\! \alpha )} u
  \right)
  \left(
\left({p\over2}\right)^{-2\alpha} +
\frac{\Gamma (1\! -\! \alpha )}{\Gamma (1\!+\!\alpha)} v
  \right)
=
\vert w\vert^2 \,.
\ee
Consider the LHS of (\ref{speLa})
as a function $F(p)$ defined on $]0,+\infty[$.
Since the RHS of (\ref{speLa})
is always non-negative some elementary analysis
gives immediately the number of solutions. The number of roots of $F(p)$
equals 0 or 1 or 2. There is no root iff $u\geq 0,v\geq 0$, and there are
two roots iff $u<0,v<0$ (it may happen that the two roots coincide giving
a multiple root). Denote by $p_1$ the smallest root of $F(p)$, if any.
In the case of two roots, let $p_2$ be the greater one. Then clearly $F(p)$
is decreasing on $]0,p_1]$ from $+\infty$ to zero and is increasing on
$[p_2,+\infty[$ from zero to the asymptotic value $uv/\alpha(1-\alpha)$.

We conclude that there are
\bea
\rm{two ~eigenvalues} &&\rm{if}~~
u<0, v<0,\mbox{ and } \det\La =uv- \al(1-\al )\vert w\vert^2 > 0 \,,
\nonumber\\
\label{sol}
\label{sol1}
\rm{no ~eigenvalue} &&\rm{if}~~
u\geq 0,v\geq 0,\mbox{ and }\det\La =uv-\al(1-\al )\vert w\vert^2\geq 0\,,
\\
\label{sol2}
\rm{one ~eigenvalue}&& \rm{otherwise} \,. \nonumber
\eea
We stress once more that all eigenvalues, if any, are negative. Again,
it may happen (when $w=0$) that two eigenvalues coincide producing
consequently a multiple eigenvalue.

For generic $\alpha$
not much can be said about what are the solutions,
except the case $w=0$ when 
$p =  2 \left( - 
\frac{u \Gamma (\alpha)}{\Gamma(2\! -\! \alpha )}
  \right)^{2-2\alpha}$ if $u<0$ and/or
$p =  2 \left( -
\frac{v \Gamma (1\! -\! \alpha )}{\Gamma (1\!+\!\alpha)}
  \right)^{2\alpha}$ if $v<0$,
and the case $u=v=0$ when $p = 2/\vert w\vert$.
An interesting particular case is $\alpha = 1/2 $ when we can give a
complete answer
about the values of the two solutions (the first case in (\ref{sol})),
\be
\label{solu}
p_{\pm} =
\frac{u\! +\! v\! \pm\left( \vert w\vert^2 + (u\! -\! v)^2\right)^{\half}}
{2(\vert w\vert^2/4 \!- \! uv)} \,,
\ee
and of the solution (the third case in (\ref{sol1})),
\be
\label{solu1}
p =
\frac{u\! +\! v\! +\!  \left( \vert w\vert^2 + (u\! -\! v)^2\right)^{\half}}
{2(\vert w\vert^2/4 \!- \! uv)} ~.
\ee
Similar, but more complicated analysis can be also performed for
$\alpha = 1/3, 1/4$ and partially for other fractional values of $\alpha$.

As far as the eigenvectors are concerned, they have to be obtained,
of course, as solutions of the differential equation (\ref{adjeigen})
including the
corresponding boundary conditions (\ref{BC}).
First of all, it is clear that in the sectors of the angular
momentum $m\neq -1, 0$,
there is a complete system of generalized (and normalized)
eigenfunctions coinciding with those of the pure A-B effect,
\be
\label{eig}
(2\pi )^{-1/2} J_{\vert m\! +\!\alpha\vert}(kr) e^{\im m\theta } ~,
~~ m= \dots  -3, -2, 1, 2 \dots , ~~k>0 ~.
\ee
Next we pass to the sectors $m=-1$ and $m=0$.

As far as the (true) eigenfunctions are concerned, the $L^2$-integrability
condition at infinity restricts the eigenvalue $k^2$ to
$k^2 <0$, and picks up a unique solution,
up to a multiplicative constant, in each sector $m= -1$ and $m=0$
(this means that both exponential growth and oscillatory behaviour
at infinity are excluded).
Hence setting as before $k=\im p$, with $p>0$,
the eigenfunction must have the form
\be
\label{xeH}
\xi H^{(1)}_{1-\alpha}(\im pr) e^{-\im \theta } +
\eta H^{(1)}_{\alpha}(\im pr)  ~,
\ee
where $\xi , \eta\in \bbc$.
The boundary conditions (\ref{BC}) lead to the following relation
between $\xi$ and $\eta$,
\be
\label{eige}
\left(
I + \left(\frac{\im p}{2}\right)^D
\Gamma (1\!-\!D)\, \La ~ \Gamma(1\!+\!D)^{-1}\,
e^{-\i\pi D} \left(\frac{\im p}{2}\right)^D \right)
\pmatrix{\xi\cr \eta} = 0 ~.
\ee
Setting the determinant of this system of linear equations to zero we
recover the equation on eigenvalues (\ref{speLa}).
Any non-trivial solution
$(\xi,\eta)$ of (\ref{eige})
determines an eigenfunction in accordance with (\ref{xeH}).

As far as the generalized eigenfunctions are concerned,
the eigenvalue equation admits a four-parameter solution
\be
\left( \xi J_{-1+\alpha}(kr)+\eta J_{1-\alpha}(kr) \right)
e^{-\im \theta } +
\xi ' J_{-\alpha}(kr)+\eta ' J_{\alpha}(kr)  ~,
\ee
where $k>0$ and $\xi , \eta, \xi ', \eta ' \in \bbc$.\\
In view of the asymptotics of
$J_{\nu}(z)\simeq
{\Gamma(1\!+\!\nu)^{-1}}(z/2)^{\nu}$ $\left(1+{\rm O}(z^2) \right)$
and again by taking into account the boundary conditions (\ref{BC})
we find the relation
\be
\label{geige}
\pmatrix{\xi\cr \xi '} = (k/2)^D \Gamma(1\! -\! D)\,
\La\, \Gamma (1\! +\! D)^{-1}\,
(k/2)^D \pmatrix{\eta\cr \eta '} \,.
\ee
A possible choice is
$$
\pmatrix{\eta\cr \eta '} =  \pmatrix{1\cr 0} ~{\rm and}~
\pmatrix{\eta\cr \eta '} =  \pmatrix{0\cr 1} $$
and in this way we obtain two independent solutions
\bea
b_1(k)&=& \left( \xi_1 J_{-1+\alpha}(kr) +
J_{1-\alpha}(kr) \right) e^{-\im \theta} +
 \eta_1  J_{-\alpha}(kr)  ~,
\nonumber\\
\label{gsol12}
b_2(k)&=& \xi_2 J_{-1+\alpha}(kr) e^{-\im \theta} +
\eta_2 J_{-\alpha}(kr)+ J_{\alpha}(kr)  ~,
\eea
where
\bea
\pmatrix{\xi_1&\xi_2\cr \eta_1&\eta_2} &=&
(k/2)^D \Gamma(1\! -\! D)\,
\La\, \Gamma (1\! +\! D)^{-1}\, (k/2)^D \nonumber\\
\label{gsol3}
&=& \left( \begin{array}{cc}
u\,\mbox{\Large$\frac{\Gamma (\alpha)}{\Gamma (2\!-\!\alpha )}
\left({k\over2}\right)$}^{2-2\alpha} &
\bar w\,\mbox{\Large${k\over 2}$} \cr
w\,\mbox{\Large${k\over 2}$} &
v\,\mbox{\Large$\frac{\Gamma (1\! -\!\alpha )}{\Gamma (1\! +\! \alpha )}
\left({k\over2}\right)$}^{2\alpha}
\end{array} \right) \ .
\eea
Set
\be
N(k)=\pmatrix{\xi_1&\xi_2\cr \eta_1&\eta_2}
+\exp(\im\pi D) \,.
\ee

Now we seek a pair of eigenfunctions which are complete and orthonormal
in the generalized sense.
In order to compute the scalar products of $b_1(k)$ and $b_2(k)$,
we need to know the integrals involving the products
$J_\mu(ay)\, J_\mu(xy)$ and $J_{-\mu}(ay)\, J_\mu(xy)$.
Recalling the relation between the functions $H^{(1)}_\mu$ and $K_\mu$
(\ref{HverK}) and using the limit value of (\ref{bat}) we have
\be
\label{batt}
\int_0^\infty H^{(1)}_\mu (ay)\, J_\mu(xy)\,y\,dy =
{1\over \im\pi a}
\left({x\over a}\right)^\mu
\left(  {1\over x\! -\! a\! -\! \im 0}-{1\over x\! +\! a} \right) \ .
\ee
Next, with the help of the distributional identity
$(x-\im 0)^{-1}={\cal P}(1/x) + \im\pi\delta (x)$,
where ${\cal P}$ denotes the principal part,
we get two identities
by taking the real and imaginary parts of (\ref{batt}),
\bea
\label{gennor}
\int_0^\infty J_\mu(ay)\, J_\mu(xy)\,y\,dy &=&
{1\over a}\, \delta (x\! -\! a) \,,\\
\int_0^\infty J_{-\mu}(ay)\, J_\mu(xy)\,y\,dy &=&
{\cos(\pi\mu)\over a}\delta (x\! -\! a) \nonumber\\
\label{gennorm}
& & + {\sin(\pi\mu)\over \pi a}\left({x\over a}\right)^\mu
\left({\cal P}({1\over x\! -\! a}) - {1\over x\! +\! a}\right)\ .
\eea

Applying (\ref{gennor})-(\ref{gennorm}) to the solutions (\ref{gsol12})
arranged in a row $B(k) = \left(b_1(k), b_2(k)\right)$,
we obtain the following $2\times 2$ matrix of scalar products,
\be
\langle B(k') ,  B(k) \rangle
= N(k')^\ast N(k) {1\over k}\, \delta (k-k') \ .
\ee
We observe that
\be
\label{detm}
\det N(k) = \left(\frac{uv}{\alpha (1\!\! -\!\!\alpha )}-w\bar w \right)
{k^2\over 4}+
 \frac{u\, \Gamma (\alpha)\, e^{\im\pi\alpha}\, k^{\!2-2\alpha}}
{2^{2\! -\! 2\alpha} ~\Gamma (2\! -\! \alpha )} 
- \frac{v\, \Gamma (1\!\! -\!\!\alpha )\, e^{-\im\pi\alpha}\, k^{\!2\alpha}}
{2^{2\alpha}~\Gamma (1\! +\! \alpha )} -1
\ee
equals minus the LHS of (\ref{detL}), with $p$
being replaced by $-\im k$,
and thus, in view of our analysis of eq. (\ref{speLa}),
$\det N(k)$ is nonvanishing for all $k\ge0$.
Therefore $\left(g_1(k), g_2(k)\right) := B(k) N(k)^{-1}$,
being given by
\bea
g_1(k;r,\theta) \!&=&\! {\det}^{\! -1}\! N(k)\,
( (\xi_1\eta_2 -\xi_2\eta_1 +\xi_1 e^{\im\pi\alpha})
J_{\alpha\!  -\! 1}(kr)e^{-\im \theta} \nonumber\\
&&
+
(\eta_2 +e^{\im\pi\alpha}) J_{1\! -\! \alpha}(kr) e^{-\im \theta}
 -\eta_1  J_{\alpha}(kr) +\eta_1 e^{\im\pi\alpha} J_{-\alpha}(kr) )  ~,
\nonumber\\
\label{ogsol12}
g_2(k;r,\theta) \!&=&\! {\det}^{\! -1}\! N(k)\,
(-\xi_2 e^{-\im\pi\alpha} J_{\alpha\!  -\! 1}(kr) e^{-\im \theta}
-\xi_2 J_{1\! -\! \alpha}(kr) e^{-\im \theta} \\
&& + (\xi_1- e^{-\im\pi\alpha}) J_{\alpha}(kr)
+(\xi_1\eta_2 -\xi_2\eta_1 -\eta_2 e^{-\im\pi\alpha}) J_{-\alpha}(kr) )~,
\nonumber
\eea
where $\xi_1$, $\xi_2$, $\eta_1$ and $\eta_2$ are defined by (\ref{gsol3}),
form a complete orthonormal basis of generalized eigenvectors
in the subspace corresponding to the absolutely continuous
spectrum in the two considered sectors,
\be
\label{genon}
\langle g_j(k') , g_{\ell}(k) \rangle =
{1\over k}\, \delta (k-k')\delta_{j,\ell} \ .
\ee

\section{Scattering}

The existence of
a complete and orthonormal basis of generalized eigenvectors
is sufficient to show that the wave (M\o ller) operators
$W_{\pm} = \lim_{t\to\pm\infty} e^{\im tH}e^{-\im tH^0}$
exist and are complete.
In fact, they can be exhibited explicitly
as well as the scattering operator $S = (W_{+})^* W_-$.

  From what we have said so far it is evident that $W_{\pm}$
and $S$ preserve the sectors $m \neq -1, 0$, and there, they
are exactly the same as in the pure A-B case;
in particular $S_m = e^{2\i\delta_m}$, where
$\delta_m = (\vert m \vert - \vert m+\alpha \vert)\pi/2$.
Thus we restrict ourselves to the subspace
$
L^2(\R_+,rdr)\otimes (\,\C\chi_{-1} \oplus \,\C\chi_{-0})\, ,
$
of remaining sectors $m = -1, 0$, which is, of course,
also preserved by all the relevant operators
(if there is no danger of confusion
we denote the restriction of an operator by the same symbol).

Using the basis $g_1(k), g_2(k)$ we define a unitary operator
${\cal F}$  from $L^2(\R_+,kdk)\otimes \C^2$ to
$L^2(\R_+,rdr)\otimes (\, \C\chi_{-1} \oplus \,\C\chi_{0})$,
by
\be
{\cal F}[\tilde \psi ] =\psi :=
\sum_{j=1}^2 \int_0^\infty g_j(k) \tilde \psi_j(k) \, k\, dk \,,
\ee
with the inverse ${\cal F}^{-1}[\psi ] = \tilde\psi  $    given by
\be
\tilde\psi_j (k) :=
\langle g_j(k) ,  \psi \rangle \ .
\ee
The operator ${\cal F}$ satisfies
\be
\label{inter}
e^{-\im tH} {\cal F} [\tilde\psi ] =
{\cal F} [e^{-\im tk^2} \, \tilde\psi]
\ee
We append the superscript `$^0$' to the relevant objects like
$g^{0}_j(k)$ or ${\cal F}^0$ corresponding to the free case
($\alpha=0$, $\Lambda = 0$); particularly
\be
g^{0}_1(k;r,\theta)=J_1(kr)\,e^{-\im\theta},\quad
g^{0}_2(k;r,\theta)=J_0(kr) \,.
\ee

Now we seek a pair of $2\times 2$ matrices
$\Omega_{\pm}=\Omega_{\pm}(k)$,
generally depending on $k$ and acting in an obvious way
as a multiplication operator on $L^2(\R_+,kdk)\otimes \C^2$,
so that
\be
\label{mol}
W_{\pm}\ {\cal F}^0 = {\cal F}\ \Omega_{\pm} \ .
\ee
Then it follows that
$S\,{\cal F}^0 = {\cal F}^0\,\Omega_{+}^{~*}\Omega_{-}$,
and this means that
$\Sigma :=\Omega_{+}^{~*}\Omega_{-}$ is nothing but the scattering matrix
in the momentum representation (restricted to the sectors $m = -1, 0$).

We have to verify that
\be
\label{limnorm}
\lim_{t\to\pm\infty} \| e^{-\im tH^0}\,{\cal F}^0[\psi]-
e^{-\im tH}\,{\cal F}[\Omega_{\pm}\psi]\| =0 \,.
\ee
Due to (\ref{inter}), the condition (\ref{limnorm}) means that
\be
\label{moll}
\lim_{t\to\pm\infty}
\sum_{j} \int_0^\infty   e^{-\im tk^2}
\vert (g^{0}_j(k) - \sum_{\ell}
g_{\ell}(k) \Omega_{\pm}(k)_{\ell j}) \psi_j(k) \vert^2
\,k\,dk= 0 \,.
\ee
It is sufficient to prove (\ref{moll})
only for 
functions 
$\psi_j(k)$ from the
dense subspace $C^\infty_0(\R_+)\subset L^2(\R_+,k\,dk)$.
By the stationary phase method, the convergence of such an integral,
as $t\to\pm\infty$, will be established provided
the coefficient standing in front of the term
$e^{\pm\i kr}$ vanishes.
In view of the known large $x$ expansion
\be
J_{\mu}(x) = (2\pi x)^{1/2} (e^{\im(x-\pi\mu/2 - \pi/4)} +
e^{-\im(x-\pi\mu/2 - \pi/4)}) +O(x^{3/4}) \,,
\ee
we obtain,
by looking separately at the coefficients in front of
$e^{-\im \theta}$ and $1$, that
\bea
&& \Omega_{+}(k)^{-1} =
\diag (-e^{-\im\pi\alpha/2},\ e^{\im\pi\alpha/2})\,
\tilde N(k)\,N(k)^{-1} \nonumber\\
&& \qquad\qquad = {\det}^{-1}N(k)\
\diag(-e^{-\im\pi\alpha/2},\ e^{\im\pi\alpha/2}) \nonumber \\
&& \!\!\!\!\!\!\!\!\!\! \times  \pmatrix{
\xi_1\eta_2 - \xi_2\eta_1  -
e^{\im\pi\alpha}(\eta_2 -\xi_1) - e^{2\im\pi\alpha} &
(e^{\im\pi\alpha} - e^{-\im\pi\alpha})\xi_2 \cr
(e^{\im\pi\alpha} - e^{-\im\pi\alpha})\eta_1 &
\xi_1\eta_2 - \xi_2\eta_1  -
e^{-\im\pi\alpha} (\eta_2 -\xi_1) - e^{-2\im\pi\alpha}
}  ,     \nonumber \\
\label{omplus}
\eea
where
\be
\tilde N(k)=\pmatrix{ \xi_1 & \xi_2 \cr \eta_1 & \eta_2 } +
\exp(-\im\pi D) \,,
\ee
and also
\be
\label{omminus}
\Omega_{-}(k) =
\mbox{diag}(-e^{-\im\pi\alpha/2},\ e^{\im\pi\alpha/2}) \,.
\ee
Note that
\be
\tilde N(k)^\ast\tilde N(k) = N(k)^\ast N(k)
\ee
and so $\Omega_+(k)$ is unitary.
Consequently, we can express (the entries of) $\Sigma$
in terms of the parameters $u, v, w$ (see (\ref{gsol3})) as
\bea
\Sigma_{11}
&=& {\det}^{-1}N(k) \times   \nonumber\\
& &\left(
e^{-\i\pi\alpha}
\left(\frac{uv}{\alpha(1\! -\!\alpha )}- |w|^2\right)
\frac{k^2}{4} +
\frac{u \Gamma(\alpha) k^{2\! -\! 2\alpha}}{2^{2\! -\! 2\alpha} \Gamma(2\! -\!\alpha )}
- \frac{v \Gamma(1\! -\!\alpha) k^{2\alpha}}{2^{2\alpha} \Gamma(1\! +\!\alpha )}
 -e^{\i\pi\alpha}  \right) \nonumber \\
\Sigma_{12}
&=& - {\det}^{-1}N(k) ~\im \,\sin(\pi\alpha )~\bar w k
\nonumber\\
\label{scat}
\Sigma_{21}
&=& - {\det}^{-1}N(k) ~\im \,\sin(\pi\alpha )~ wk \\
\Sigma_{22}
&=& {\det}^{-1}N(k) \times \nonumber\\
& &\left(
e^{\im\pi\alpha} \left(\frac{uv}{\alpha(1\! -\!\alpha )}-|w|^2 \right)
\frac{k^2}{4} +
\frac{u \Gamma(\alpha) k^{2\! -\! 2\alpha}}{2^{2\! -\! 2\alpha}\Gamma(2\! -\!\alpha )}
- \frac{v \Gamma(1\! -\!\alpha )k^{2\alpha} }{2^{2\alpha}\Gamma(1\! +\!\alpha )}
- e^{-\im\pi\alpha} \right)
\ , \nonumber
\eea
where ${\det}^{-1}N(k)$ is given by (\ref{detm}).
Obviously, $\Sigma$ is unitary.

Let us specialize these formulae to three particular cases.

\noindent(i) If $w=0$ (conserved angular momentum) then
\be
\Sigma=\diag \left(
\frac{ u\Gamma(\alpha)(k/2)^{2-2\alpha}-
e^{\im\pi\alpha}{\Gamma(2\! -\!\alpha )} }
{e^{\im\pi\alpha} u\Gamma(\alpha)(k/2)^{2-2\alpha}-
{\Gamma(2\! -\!\alpha )} },\
\frac{ v\Gamma(1\!-\!\alpha)(k/2)^{2\alpha}+
e^{-\im\pi\alpha}{\Gamma(1\! +\!\alpha )} }
{e^{-\im\pi\alpha} v\Gamma(1\!-\!\alpha)(k/2)^{2\alpha}+
{\Gamma(1\! +\!\alpha )} }
\right) \,.
\ee

\noindent(ii) If $u=v=0$ (maximal nonconservation of angular momentum)
then
\be
\label{scatuv}
\Sigma =
Q^{-1}\times \pmatrix{
e^{-\im\pi\alpha} |w|^2 (k^2/4) + e^{\im\pi\alpha}
&
\im\sin(\pi\alpha)~\bar w k
\cr
\im\sin(\pi\alpha)~ w k
&
e^{\im\pi\alpha} |w|^2 (k^2/4) + e^{-\im\pi\alpha}
}       \ ,
\ee
where
$$
Q= |w|^2 {k^2\over 4} +1 \ .
$$

\noindent (iii) If $\alpha = 1/2$ then
\be
\label{scata}
\Sigma =
Q^{-1}\times \pmatrix{
-\im - \im(uv\! -\! |w|^2/4)k^2 + (u\! -\! v)k
& -\im \bar w k \cr
-\im w k &
\im + \im(uv\! -\! |w|^2/4)k^2 + (u\! -\! v)k
}       \ ,
\ee
where
$$
Q=-1+(uv-|w|^2/4)k^2 + \im (u+v)k \, .
$$

We conclude this section by giving (the kernel of) the full
scattering operator in the angular representation,
\be
S(k;\theta , \theta_0 ) = {1\over 2\pi}
\sum_{m,n=-\infty}^{\infty} S(k;m,n)e^{\im(m\theta - n \theta_0)} \ ,
\ee
where
$$
S(k;m,n) = \left\{
\begin{array}{ll}
e^{\im\pi (\vert m\vert-\vert m+\alpha \vert)}\,
\delta_{mn} ~&\mbox{if}~m, n\neq -1, 0
\nonumber\\
\Sigma_{m+2,n+2}  ~& \mbox{if}~m, n = -1, 0
\end{array}
\right.
$$
(the shift by 2 is due to our labelling of rows and columns of $\Sigma$).
The double infinite sum can be performed in the sectors $m, n\neq -1, 0$
(a known result borrowed from the pure A-B effect) and yields
\bea
S(k;\theta , \theta_0 ) &=&
\cos(\pi\alpha) \delta (\theta\! -\!\theta_0 ) + {1\over 2\pi}
\sin(\pi\alpha) \frac{ e^{-\im(\theta - \theta_0)/2} }
{\sin (\theta /2 - \theta_0 /2)}
\nonumber \\
\label{scatt}
&+& {1\over 2\pi} \sum_{m,n=-1}^{0}
\left( \Sigma_{m+2,n+2} - e^{-(2m+1)\im\pi\alpha}\delta_{mn} \right)
e^{\im(m\theta - n\theta_0)} \ .
\eea
We recall that the differential cross-section in the plane is given by
$d\sigma(\theta)/d\theta =
(2\pi/k)\vert S(k;\theta ,\theta_0)\vert^2 $.

\section{Conclusions}

We have introduced and studied a five-parameter family of operators which
describe
a quantum mechanical particle interacting with a magnetic
flux $\alpha$ caused by an infinite thin material solenoid.
We conjecture that in some well defined limit
(when the thickness of the solenoid $\to 0$ and its length $\to \infty$)
which should be largely independent
on the details of approximating potentials
(and which however goes beyond the scope of this paper)
such a situation is described by a (singular) potential barrier
and by a electromagnetic potential concentrated along the $z$ axis
(the magnetic field vanishes in the remaining region).
One of the five parameters is just the value of the flux $\alpha$
and the other four correspond to the strength of a singular potential barrier
(sort of a combination of Dirac $\delta$ and $\delta^\prime$) and
can be interpreted as penetrability coefficients of the shielded solenoid.

A general operator of this family corresponds to
an intricate mixture between
the Aharonov-Bohm effect and the point interactions which
is manifested more concretely via the mixing between the angular
and the radial boundary conditions.
It is interesting that the result we have obtained is richer than
a simple superposition of the point interaction and the pure A-B effect.
For instance, for a range of parameters there are two bound states
possible while for the usual interactions with a support
concentrated along the $z$ axis and symmetric under the $z$-translations
(or, equivalently, for a point interaction in two dimensions)
there is always (except of the free case) exactly one bound state,
and for the pure A-B effect there are no bound states at all.

In the present paper we have derived an explicit
formula for the scattering
matrix $S(k;\theta,\theta_0)$ depending on the five parameters.
Naturally, it would be of interest to examine the differential
cross-section numerically, particularly its dependance on the
parameters, and to deduce some physical consequences. This is what we
plan to do separately.

It is a matter of experimental measurements (interference, scattering or 
in condensed matter) to establish which of the extensions in our family 
correspond to realizable models.
On general grounds, one can distinguish some class of extensions 
by eg. symmetry requirements (conservation of the angular momentum),
or postulating that there are no bound states
(solenoid as a repulsive barrier).

After completing most of our work we became aware of 
\cit{RA} which has an overlap with some of our results.
We have been also informed that a related preprint 
by R. Adami and A. Teta is in preparation.

\vskip 24pt
{\bf Acknowledgements.}\
P.S. wishes to  gratefully acknowledge the partial
support from Grant No. 201/94/0708 of Czech GA.

\end{document}